# What is Stochastic Supervenience?

*Youheng Zhang*

https://orcid.org/0000-0002-2363-487X

*School of Marxism, University of Electronic Science and Technology of China, Chengdu, Sichuan, China*

zhangyouheng@whut.edu.cn

**Abstract:** Standard formulations of supervenience typically construe the determinandum as a point-value strictly fixed by base-level states. Yet, in scientific domains ranging from statistical mechanics to deep learning, basal structures frequently determine law-governed families of probability measures rather than single outcomes. This paper develops a general framework for stochastic supervenience, representing this dependence via Markov kernels that map base states to higher-level distributions. I formulate axioms securing law-like fixation, non-degeneracy, and directional asymmetry, demonstrating that classical deterministic supervenience is recovered as the Dirac boundary case of this broader topological space. To render these metaphysical commitments empirically tractable, the paper integrates information-theoretic diagnostics—such as normalized mutual information, divergence spectra, and tail sensitivity. These indices serve to distinguish genuine structural stochasticity from epistemic noise, stratify degrees of distributional multiple realization, and diagnose intervention-salient macro-organization. The resulting framework offers a conservative extension of physicalist dependence, reconciling the priority of the base level with the structured uncertainty ubiquitous in the special sciences.

## 1 Introduction

Classical discussions of supervenience standardly proceed within a familiar tripartite schema. Following Kim (1990), supervenience involves covariance, dependence, and non-reducibility. Weak supervenience states that, within a single possible world, identity of the base properties guarantees identity of the higher-level properties. Strong and global supervenience reinforce modal stability and determinacy, respectively, by securing cross-world property matching and by requiring an isomorphism in the overall distribution of properties across worlds (Kim 1984; 1987). Mainstream discussions remain within this schematic framework, concentrating on its interpretation and application (e.g., Chen 2011; Hoffmann and Newen 2007; Kovacs 2019; Leuenberger 2009; Moyer 2008; Napoletano 2015; Shagrir 2002; 2013). Classical theories further share a background presupposition: what is determined—the determinandum— is a value (or set of values) of higher-level properties.

When we confront cases in which outputs are systematically indeterministic yet plainly tightly constrained by underlying structure and law, this framework proves rigid. Faced with phenomena that are probabilistic in a structured way but not mere noise, one is pushed either (i) to retain determinism (and thereby lose a stratified layer of uncertainty) or (ii) to embrace strong forms of irreducibility or emergence. Yet in many central scientific domains the underlying structures and laws do not appear to fix a single point value outcome, but rather to fix a stable family of probability distributions. Examples include:

- the Born probabilities of quantum measurement outcomes (see Born 1955; Wigner 1963; Brukner 2017);
- power laws and species abundance distributions in ecological and social complex systems (Hubbell 2001, 32-45; Newman 2005; Bettencourt et al. 2007);
- the direct modeling of conditional and generative distributions in statistical learning and deep generative architectures (Vapnik 1999; Gal and Ghahramani 2016; Kingma et al. 2014).

These cases suggest that if we continue to insist on pointwise functional determination, we cannot distinguish between (1) an ontic, law-governed probabilistic structure and (2) residual epistemic ignorance.

Work in the philosophy of physics has developed sophisticated accounts of how such probability distributions arise from underlying dynamics. Myrvold (2021) offers a hybrid theory of probabilities, treating Gibbsian and Boltzmannian probability measures as central theoretical posits that mediate between micro-dynamics and macroscopic regularities. Wallace and Frigg (2021), as well as Wallace (2019), analyze Gibbsian statistical mechanics in detail, showing how ensemble measures and invariant distributions can underwrite experimental practice and connect microscopic and thermodynamic descriptions. These approaches elucidate the origin and empirical role of statistical-mechanical probabilities, but they do not explicitly formulate the dependence between base-level states and higher-level probabilistic structures in supervenience-style terms. The framework of stochastic supervenience proposed here is meant to complement such accounts by articulating that dependence as a law-like mapping from base states to families of higher-level probability measures.

What calls for revision, then, is not the supervenience schema per se but the nature of its determinandum: from point-values to distributional structures. Kim himself, in passing remarks across different texts, explicitly entertains a "probabilistic" or "stochastic" supervenience:

> "I believe there may well be a viable concept of statistical or stochastic emergence, which assigns a stable objective chance of the emergence of a property given that an appropriate basal condition is present. But stochastic emergence in this sense must be based on statistical laws with nomological force, and these laws can warrant talk of stochastic supervenience and stochastic necessitation." (Kim 2006, p. 550)

"The idea of probabilistic supervenience is clearly coherent and deserves consideration. But we will presumably need stable lawlike probabilities grounded in the laws at the basal level. Details of this approach need to be worked out. I hope it is obvious that the condition that the basal conditions are merely 'necessary' for the occurrence of an emergent property will not do." (Kim 2013, p. 58)

Although recent years have seen some discussion of stochastic or probabilistic supervenience (e.g., Craver 2017; Gebharter and Sekatskaya 2024; Moorfoot 2024), this line of inquiry has not been systematically developed, nor has it been integrated into the main structure of debates about the relations between fundamental and higher-level descriptions in science more generally. Meanwhile, familiar issues—such as multiple realizability, the status of macro-models, and the proper understanding of "autonomy" and "emergence"—continue to encourage a binary choice between "thoroughgoing physical determinism" and "strong emergence" or "dual-track" ontologies (e.g., Sober 1999; Polger 2015; Baysan 2019; Stenwall 2021; Kroedel 2015; Frankish 2007; Hendricks 2024; Goff 2017).

Doob (1942, p. 648) notes that "A stochastic process is simply a probability process; that is, any process in nature whose evolution we can analyze successfully in terms of probability." Thus, "stochastic" refers to objects or processes whose evolution exhibits uncertainty and randomness, while "probabilistic" refers to the use of probabilistic methods to model, analyze, and understand such phenomena. Although Kim used these two terms interchangeably, I adopt "stochastic" here to avoid the risk that "probabilistic" might, in some philosophical contexts, be interpreted epistemologically or as mere frequency aggregation, rather than indicating nomological, law-governed objective probabilities.

Throughout I adopt an objective-chance (law-governed) reading of probabilities: they are nomological constraints (chances) rather than merely epistemic degrees of belief or frequency summaries.

This avoids collapsing stochastic supervenience into a purely pragmatic modeling convenience and aligns it with chance-based interpretations of physical modality.

Building on Kim's suggestions and existing scientific practice, this paper develops a framework of stochastic supervenience aimed at articulating law-like dependence in settings of structured uncertainty. The core move is to shift the determined object from a specific instantiation of higher-level properties to a probability distribution over those properties. In this way, we preserve physicalist priority and law-like constraints at the basal level while allowing higher-level descriptions a form of autonomy in terms of their distributional patterns.

The significance of stochastic supervenience lies not in weakening the concept of supervenience, but in redefining and expanding its scope. It does not introduce arbitrary noise; rather, basal states lawfully determine a stable cluster (or family) of probability measures. Nor does it deny the asymmetry inherent in supervenience; instead, it embeds this asymmetry within a hierarchy from fixed values to distributions. Classical deterministic supervenience appears as a boundary case—the Dirac limit—of a more general space of base-to-distribution kernels.

Information-theoretic diagnostics and intervention-based measures provide a way to render this probabilistic dependence empirically testable. They help distinguish genuinely structural stochastic patterns from mere noise or trivial relabelings, and they quantify the extent to which higher-level descriptions contribute non-redundant information and intervention-sensitive structure.

The plan is as follows. Section 2 introduces the unified formal framework: section 2.1 supplies the measure-theoretic and information-theoretic toolkit; section 2.2 axiomatizes stochastic supervenience and embeds classical supervenience as a Dirac boundary case; sections 2.3–2.5 integrate distributional realization with informational indices to formulate criteria that distinguish law-locked probabilistic structure from mere noise or trivial compression, and establish propositions concerning the deterministic limit, preservation under coarse-graining and composition, and intervention-level discriminability. Section 3 articulates the broader philosophical payoffs for understanding inter-theoretic dependence, autonomy, and emergence in a probabilistic setting, and defends the framework against several natural objections.

**2 Unified Framework**

**2.1 Formal Preliminaries**

Let the space of basal (base-level) states be a measurable space $(B, \Sigma_B)$, and the (higher-level) property space be $(A, \Sigma_A)$. Write $\Delta(A)$ for the set of all probability measures on $A$. Fix a background set of laws $L$ (dynamical or bridge principles). Call a structural feature law-like if it is invariant across all possible worlds compatible with $L$; let $W_L$ denote the set of such worlds.

Stochastic supervenience is represented by a (Markov) kernel $\Phi: B \to \Delta(A)$: given $b \in B$, the laws assign a higher-level probability measure $\Phi(b)$. If for every b, $\Phi(b)$ is Dirac, then there is a unique function $f: B \to A$ with $\Phi(b) = \delta_{\{f(b)\}}$, and we collapse back to classical deterministic supervenience. In the general case, what is fixed is a distributional profile rather than a single point-value outcome.

To compare higher-level distributions induced by distinct basal states or by their aggregates, choose a distance $d$ on $\Delta(A)$ satisfying only (1) separation and (2) the triangle inequality, so as not to hard-wire a specific geometry into later arguments (see Gibbs and Su 2002). In discrete settings one may use the Jensen–Shannon distance $d_{JS}$ (Lin 1991); if A carries a natural geometry, a p-Wasserstein distance $W_p$ can reflect "mass transport cost" (Villani 2009); when sensitivity to small low-probability differences is required, one may locally invoke the total variation distance $d_{TV}$ (Le Cam and Yang 2000). With $d_{JS}$, choosing logarithm base 2 yields $d_{JS} \in [0,1]$, facilitating later similarity or normalization measures (Lin 1991).

Information-theoretic notation is standard: $H(A)$ is the Shannon entropy of the higher-level variable; mutual information $I(B; A) = H(A) - H(A \mid B)$ (Cover and Thomas 2005a; Shannon 1948). Define the normalized quantity: $A^* := I(B; A) / H(A)$. $A^*$ measures the extent to which the base $B$ removes $A$'s original uncertainty (Vinh et al. 2010). Values $A^* \approx 1$ indicate the higher-level is almost distributionally fixed by the base; $0 < A^* < 1$ indicates a law-like relation that nonetheless preserves genuine probabilistic structure; $A^* = 0$ means the base has no discriminatory power over the "shape" of A's distribution.

For macro-level analysis we group basal states into blocks $B_i$ (a coarse-graining). Given within-block weights $w(b)$ (e.g. empirical frequencies, priors, or uniform), define the block-averaged measure

$$\Phi_i = \Sigma_{\{b \in B_i\}} w(b) \, \Phi(b).$$

Thus $\Phi_i$ is the representative (weighted) measure of block $B_i$. A common situation: two aggregates $B_i$,

$B_j$ yield representative distributions $\Phi_i$, $\Phi_j$ that are close over their main (high-probability) bodies yet diverge markedly in low-probability tails (Clauset et al. 2009). To extract tail structure explicitly, fix a small threshold $p_0$ (e.g. 0.01 - 0.05) and define the shared tail set:

$$S_{\{ij\}} := \{ a \in A : \Phi_i(a) < p_0 \text{ and } \Phi_j(a) < p_0 \}.$$

Let $Z_i := \Phi_i(S_{\{ij\}})$, $Z_j := \Phi_j(S_{\{ij\}})$. Define tail-normalized distributions (for $a \in S_{\{ij\}}$):

$$\Phi_{i\{tail,(ij)\}}(a) := \Phi_i(a) / Z_i, \quad \Phi_{j\{tail,(ij)\}}(a) := \Phi_j(a) / Z_j,$$

and set them to 0 outside $S_{\{ij\}}$. If $Z_i = 0$ or $Z_j = 0$ (no shared tail mass), set $TailDiv_{\{ij\}} = 0$. Tail divergence is then

$$TailDiv_{\{ij\}} := d_{TV}( \Phi_{i\{tail,(ij)\}}, \backslash \Phi_{j\{tail,(ij)\}} ).$$

$TailDiv_{\{ij\}}$ captures shape discriminability within events both aggregates regard as rare: larger values mean that even in the jointly low-probability region the structures differ substantially. This decouples similarity in the main body $A \setminus S_{\{ij\}}$ from structural divergence in the shared tail; later, in classifying distributional realization levels (robust / fragile / pseudo-equivalent), I will use body distances together with $TailDiv_{\{ij\}}$ as classification criteria.

To assess whether a macro aggregation yields causal or informational gain, I use Effective Information (EI). Here EI refers to mutual information under uniform intervention over the cause variable (Pearl's do-semantics; each realizable state is set with equal probability) (Pearl 2009; Cover and Thomas 2005a; Hoel et al. 2013):

$$EI = H(Effect) - H(Effect \mid Cause) = I(Cause; Effect).$$

Let the micro cause variable be $B$ and the macro cause variable be $G(B)$, where $G$ is a partition map. Macro-level interventions take each block of $G(B)$ equiprobably and are uniform over micro states within a block (Hoel et al. 2013; Hoel 2017). Define the increment:

$$\Delta EI = EI_{macro} - EI_{micro} = EI(G(B) \to A) - EI(B \to A).$$

$\Delta EI > 0$ indicates that at the intervenable macro scale the higher-level response structure is sharpened rather than merely information-losing; equivalently, $\Delta EI > 0 \Leftrightarrow [H(B \mid A) - H(G(B) \mid A)] > [H(B) - H(G(B))]$: the reduction in conditional uncertainty / degeneracy induced by aggregation exceeds the entropy loss from fewer input states. With multiple input variables, a Partial Information Decomposition (PID) can split $I(Inputs; A)$ into redundant, unique, and synergistic components; if synergy increases post-aggregation, certain probabilistic constraints emerge only at the aggregated level (not in isolated micro components), supporting the non-triviality of macro "new" structure (Williams and Beer 2010; Mages et al. 2024).

To capture heterogeneity within the base layer, we may also use the overall distributional divergence spectrum $(D_{set})$[1]: the collection $\{d_{JS}(\Phi(b), \Phi(b')) : b, b' \in B\}$ together with summary statistics (mean, quantiles, range) (Lin 1991; Rao 1982). This spectrum is used alongside $A^*$: the former exhibits the shape of residual dispersion, the latter gives the proportion explained. If $A^*$ is moderate while the divergence spectrum collapses sharply over the main body, this often signals that structural randomness concentrates in tail patterns (see Coles 2001; Clauset et al. 2009)—the motivating scenario for introducing $TailDiv$.

In summary: $\Phi$ supplies base-to-higher-level probabilistic assignment; $d$ and its derived spectrum characterize distributional dissimilarity; $A^*$ quantifies the degree of law-fixation; $\Phi_i$ and $TailDiv_{\{ij\}}$ separate body and tail structure; $\Delta EI$ and synergy components test whether a macro scale yields additional intervenable information or distributional constraints. Section 2.2 will formally define stochastic supervenience within this framework and exhibit the embedding of classical supervenience as a limiting case: every deterministic supervenience function $f: B \to A$ corresponds to $\Phi(b) = \delta_{\{f(b)\}}$, and no deterministic result is lost—only a strictly finer classification of non-deterministic law-locked structure is added.

**2.2 Axioms of Stochastic Supervenience**

In 2.1 we have fixed: the base state space $B$, the macro (higher-level) property space $A$, the map $\Phi: B \to \Delta(A)$, the background set of laws $L$, and several distance/information measures. The axioms below concern only $\Phi$ and $L$; the objects $A^*, \Delta EI, TailDiv$, etc. are analytic evaluative instruments and do not themselves constitute supervenience criteria.

**Definition 2.1** $A$ is said to stochastically supervene on $B$ if there exists a map $\Phi: B \to \Delta(A)$ such that:

(SS1) Law-Like Fixation. The same base state $b$ induces the same probability measure $\Phi(b)$ in every possible world compatible with $L$; that is, for all $w \in W_L$ and all $b \in B$, $P_w(A \in \cdot \mid B = b) = \Phi(b)$ [2].

---

[1] The "distributional divergence spectrum" is a term specific to this paper, denoting the set of all pairwise distances between conditional distributions together with their summaries. The rationale follows diversity formulations based on pairwise dissimilarities (see Rao 1982) and is used to make explicit the information lost when one reports only a single averaged divergence.

(SS2) Ontic Non-Degeneracy. There exist $b_1 \neq b_2$ with $\Phi(b_1) \neq \Phi(b_2)$.

(SS3) Directional Asymmetry. If there exists b with $\Phi(b)$ non-Dirac, then there is no $L$-admissible (measurable, law-preserving) bijection $R: \Phi(B) \to B$ such that $\Phi \circ R = id_{\{\Phi(B)\}}$ and $R \circ \Phi = id_B$ [3].

(SS4) Epistemic Closure Criterion. For any admissible refinement of base predicates—i.e. any partition or predicate enrichment definable within the original law set $L$ that (1) introduces no novel measurable structure beyond $L$'s discriminatory resources; (2) does not alter the underlying base state $b$—if after such refinement $\Phi(b)$ remains non-Dirac (and chosen divergence indices remain strictly positive), then the residual probabilistic structure is treated as ontic rather than epistemic [4].

(SS5) Distributional Necessitation. Write $b_1 \sim_L b_2$ to mean that $b_1$ and $b_2$ are law-and-observationally indiscernible across all worlds compatible with $L$. If $b_1 \sim_L b_2$, then $\Phi(b_1) = \Phi(b_2)$ [5].

**Definition 2.2** If $\Phi$ satisfies (SS1), (SS3), (SS4), and (SS5) and, moreover, $\Phi(b) = \delta_{\{f(b)\}}$ (a Dirac measure) for some function $f: B \to A$ and all $b \in B$, we recover deterministic supervenience. In that case:

---

[2] The identity $P_w(A \in \cdot \mid B = b) = \Phi(b)$ asserts equality of conditional probability measures: it states that the function sending each measurable $E$ to $P_w(A \in E \mid B = b)$ coincides with $\Phi(b)$. The "·" is a placeholder for events. (Regular conditional probabilities are assumed where needed.)

[3] Rationale for conditional directional asymmetry: were there, in a genuinely stochastic (some non-Dirac $\Phi(b)$) setting, a law-like bijection $R$ under $L$ rendering $\Phi$ and $B$ bijectively isomorphic, the higher-level probabilistic structure would amount to mere relabeling of base states, erasing the intended direction "higher-level determined (probabilistically) by base" rather than mutual isomorphism. We therefore exclude such strict law-like isomorphism only in non-Dirac cases. One-sided (e.g. existence of a section) or local invertible fragments are not excluded; they do not undermine global directionality. In the fully deterministic limit where all $\Phi(b)$ are Dirac, a bijective relabeling may exist; this is treated as a degenerate boundary and (SS3) becomes vacuous.

[4] Formal refinement exclusion. A non-Dirac $\Phi(b)$ cannot be represented purely as ignorance over a finer $L$-silent partition $\pi: B \to B$ with a deterministic $f: B \to A$ and fibre measures $\rho_b$ (i.e. $\Phi(b) = f_*\rho_b$) unless new $L$-recognizable predicates are introduced. Otherwise randomness would be epistemic only.

[5] L-indiscernibility (($\sim_L$): $b_1 \sim_L b_2$ iff for every law-like base predicate $P$ permitted by $L$ (excluding predicates whose definition depends on $\Phi$), $P(b_1) = P(b_2)$. Thus $\Phi$ depends only on the $L$ equivalence class of $b$ and ignores law-redundant individual differences.

(1) If f is non-constant, (SS2) still holds (there exist $b_1 \neq b_2$ with $f(b_1) \neq f(b_2)$;

(2) If f is constant, the system is fully degenerate, higher-level properties lack discriminability, and (SS2) fails—this case is structurally trivial.

Condition (SS4) becomes vacuous in the all-Dirac limit. Since (SS3) is conditional, it is automatically vacuous when all $\Phi(b)$ are Dirac (including bijective $f$). This establishes stochastic supervenience as a conservative extension: classical supervenience occupies the Dirac boundary of the $\Phi$-space (as illustrated in **Fig. 1**).

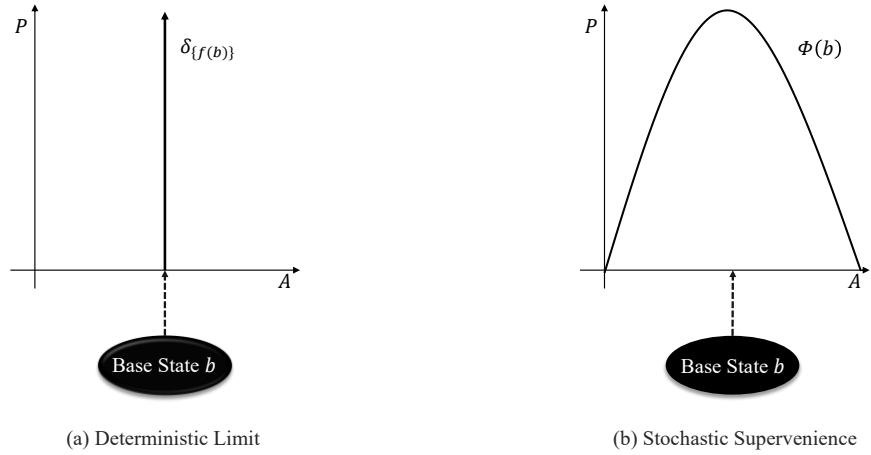

**Fig. 1** Stochastic supervenience as a conservative extension. (a) Classical supervenience appears as the boundary case where the distribution collapses to a Dirac measure $\delta_{\{f(b)\}}$. (b) The general case where the base state $b$ fixes a probability profile $\Phi(b)$ rather than a single value.

## 2.3 Distributional Realization

Building on 2.2 which introduced stochastic supervenience and the tools of section 2.1, we now have: (1) candidate realization clusters $B_i$ with their representative measure $\Phi_i = \Sigma_{\{b \in B_i\}} w_i(b) \Phi(b)$; (2) an overall distributional distance structure via the J–S distance $d_{JS}$; (3) the shared tail set $S_{\{ij\}}$ and the tail-divergence index $TailDiv_{\{ij\}}$ (see **2.1**). This subsection introduces no new axioms of stochastic supervenience; instead, it provides an empirically oriented, graded description of the distributional stability of multiple realizations using these existing metrics.

In the deterministic setting, multiple realizability means that several distinct sets of micro-level structures can yield the same higher-level property $a$ (as a single point value)[6]. Under the stochastic supervenience framework, the same macro property $a$ is no longer a point, but is characterized by a law-governed family of probability measures on $A$: different basal states $b$ map via $\Phi(b)$ to output distributions that may be extremely close overall, or may diverge only in low-probability regions or under specific conditional scenarios. The two illustrative cases in **Appendix A (A.1: Neural Networks; A.2: Multi-cluster Markov system)** exemplify two salient patterns (plotted in **Fig.2**). Accordingly, we focus on certain sets of basal states (realization clusters) $B_i$ regarded as candidate joint realizers of the same macro property. When their differences are confined chiefly to tails or higher-order shape while their high-probability "body" region is sufficiently similar, graded statuses of robustness, fragility, or pseudo-equivalence can emerge[7] (the semantics of $B_i$ and weighting choices are given in Footnote 6).

Given a realization cluster $B_i$, choose weights $w_i(b) \geq 0$ with $\Sigma_{\{b \in B_i\}} w_i(b) = 1$ (replace the sum by an integral for the continuous case) and form the representative measure $\Phi_i = \Sigma_{\{b \in B_i\}} w_i(b) \Phi(b)$. This representative is a convex combination of the $\Phi(b)$ within the cluster; it facilitates comparison but may smooth over internal multimodality (see **Footnote 6**). Define overall similarity:

$$Sim(B_i, B_j) = 1 - d_{JS}(\Phi_i, \Phi_j),$$

where $d_{JS}$ is the J–S distance with logarithm base 2, so $Sim \in [0,1]$. Fix thresholds $1 \geq \tau_{high} > \tau_{low} \geq 0$. Recalling Section 2.1, define the shared tail set:

---

[6] I use "realization" in the dimensioned, micro-macro sense discussed by Gillett (2002) and Polger and Shapiro (2008): lower-level or microstructural properties of a system realize higher-level properties of that same system (for example, lattice-level properties realizing the hardness of a diamond). This contrasts with the familiar "flat" use of realization, on which heterogeneous systems that do the same thing in different ways – for instance, waiter's corkscrews and lever-type corkscrews – count as multiple realizers of a single functional kind (cf. Bennett, 2011; Polger & Shapiro, 2008). The framework of stochastic supervenience is meant to analyze the first, dimensioned sort of realization, where law-governed probability distributions link microstates to macro-profiles. Whether an analogous treatment extends to flat realization is a further question that I set aside here.

[7] Realization clusters and representative measures: A realization cluster $B_i$ is a set of basal states treated as jointly realizing the same higher-level property a (it may arise from unions of $\sim_L$ equivalence classes or from empirical clustering). The weights $w_i(b)$ may be empirical frequencies, prior assessments, or uniform; in the continuous case use $\int w_i(b) \Phi(b) \, db$. The convex mixture $\Phi_i$ can understate internal multimodal differences; if structural preservation is required, one may instead use a J–S barycenter minimizing $\Sigma\{b \in B_i\} w_i(b) d_{JS}(\Phi(b), \Psi)^2$, or replace $\Phi_i$ by a Wasserstein barycenter.

$$S_{\{ij\}} = \{\, a \in A \mid \Phi_i(a) < p_0 \text{ and } \Phi_j(a) < p_0 \,\}, \text{ with } 0 < p_0 < 1.$$

If $\Phi_i(S_{\{ij\}}) > 0$ and $\Phi_j(S_{\{ij\}}) > 0$, then for any measurable $E \subseteq A$ define tail-normalized measures: $\Phi_{i\{tail,ij\}}(E) = \Phi_i(E \cap S_{\{ij\}}) / \Phi_i(S_{\{ij\}})$, $\Phi_{j\{tail,ij\}}(E) = \Phi_j(E \cap S_{\{ij\}}) / \Phi_j(S_{\{ij\}})$, and set $TailDiv_{\{ij\}} = d_{TV}(\Phi_{i\{tail,ij\}}, \Phi_{j\{tail,ij\}})$ (where $d_{TV}$ see **2.1**). Otherwise let $TailDiv_{\{ij\}} = 0$. (Below, write $TailDiv$ for brevity) Choose $0 \leq \theta_{tail} < \theta_{pseudo} \leq 1$, and classify **(visualized in Fig. 2)**:

(1) Robust: $Sim \geq \tau_{high}$ and $TailDiv \leq \theta_{tail}$;

(2) Fragile: $\tau_{low} \leq Sim < \tau_{high}$ or $Sim \geq \tau_{high}$ and $\theta_{tail} < TailDiv \leq \theta_{pseudo}$;

(3) Pseudo-equivalent: $Sim \geq \tau_{high}$ and $TailDiv > \theta_{pseudo}$ or $Sim < \tau_{low}$.

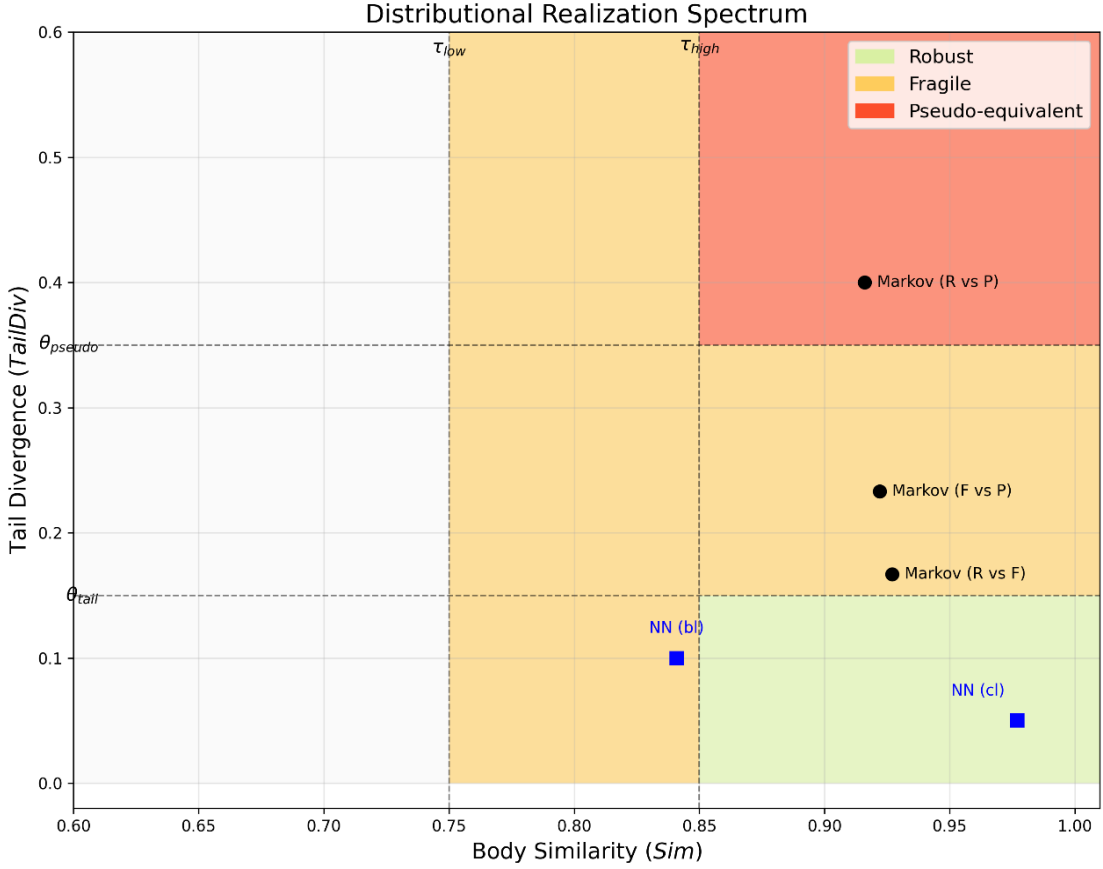

**Fig. 2** The spectrum of distributional realization. The classification phase space is defined by body similarity $Sim$ and tail divergence $TailDiv$. Shaded regions correspond to the structural stability categories defined in Section 2.3: robust (green), fragile (yellow), and pseudo-equivalent (red). The decision boundaries are determined by the thresholds $\tau_{\{high/low\}}$ and $\theta_{\{tail/pseudo\}}$. Markers indicate the empirical cases analyzed in **Appendix A** (squares: Neural Networks; circles: Multi-cluster Markov System), illustrating how the metrics discriminate between functionally distinct realizations.

If $TailDiv$ has not yet been computed, use the provisional rule: $Sim \geq \tau_{high}$ → robust; $\tau_{low} \leq Sim < \tau_{high}$ → fragile; $Sim < \tau_{low}$ → pseudo-equivalent. This stratification is purely a methodological annotation of distributional proximity among realization clusters under a fixed Φ. It differentiates: (1) robust cases where body and tail both align; (2) fragile cases where bodies are similar but the structure is sensitive to rare perturbations; and (3) pseudo-equivalent cases where bodies appear similar yet the tail harbors significant divergences. The scheme interfaces with $A^*$, $\Delta EI$, and the divergence spectrum from Section 2.1: when $A^*$ is moderate, identifying pseudo-equivalent clusters helps locate refinements that might raise $\Delta EI$; for clusters already classified as robust, further refinement typically yields no $\Delta EI$ gain.

**2.4 Autonomy Diagnostics**

This section consolidates previously introduced analytic indices for later assessment of distributed autonomy. They provide a structural profile of a fixed $\Phi$; technical details appear in **Appendix B**.

(1) $A^* = I(B; A) / H(A)$. A value $0 < A^* < 1$ indicates that the basal level lawfully constrains, yet does not collapse, the macro-level probabilistic structure; $A^* = 1$ marks the deterministic limit; $A^* = 0$ would violate (SS2) (estimation details: **Appendix B.1**).

(2) Distributional divergence spectrum $D_{set} = \{d_{JS}(\Phi(b), \Phi(b'))\}$. Report statistical summaries (mean, quantiles, range). Comparison to a permutation baseline (see **2.1** and **Appendix B.2**, **B.4**) helps exclude spurious "micro-noise camouflage."

(3) Effective information increment $\Delta EI = EI(G(B) \to A) - EI(B \to A)$. $\Delta EI > 0$ indicates that the macro-level coarse-graining sharpens intervention distinguishability (see **2.1** and **Appendix B.3**; intervention legitimacy criteria: **B.7**).

(4) Synergy/redundancy structure (Standard PID methods; not expanded in Appendix B). An increase in the synergy share after aggregation ($\Delta Synergy > 0$) shows that constraints emerge in integrated form at the macro scale (supporting non-trivial organization) and provides mechanistic background for the intervention differences stated in **Proposition 2.8**.

(5) Tail sensitivity ($TailDiv$). For realization cluster pairs with high body similarity ($Sim \geq \tau_{high}$), examine divergence on their shared tail ($TailDiv$, or a weighted summary thereof) to capture latent fragility or pseudo-equivalence in low-probability patterns (see **2.1** and **Appendix B.2**; TS: **B.8**).

Optional diagnostics: MDL-based model selection (**B.5**) and cross-context robustness (**B.6**).

## 2.5 Propositions and Proofs

This subsection records several key propositions linking the qualitative axioms of stochastic supervenience with the quantitative indices of 2.1–2.4. Proof sketches are given where they illuminate the structure; full technical details are relegated to the appendices.

**Proposition 2.3** The following three conditions are equivalent: (1) For all $b \in B$, $\Phi(b)$ is a Dirac measure; (2) $H(A \mid B) = 0$; (3) $A^* = 1$.

*Proof* The standard mutual information identity gives $A^* = (H(A) - H(A|B))/H(A)$. Hence (I) ⇒ (II); (II) ⇒ (III); (III) ⇒ (II); (II) ⇒ (I) (zero conditional entropy implies each conditional distribution is Dirac). If all $\Phi(b) = \delta_{\{a_0\}}$ (a constant Dirac), then $H(A) = 0$ and $I(B; A) = 0$, making $A^*$ formally $0/0$; this fully degenerate case is excluded from the (I)–(III) equivalence discussion.

**Proposition 2.4** If (SS2) holds and $H(A) > 0$, then $0 < A^* \leq 1$. Conversely, if $A^* = 0$, then $I(B; A) = 0 \Rightarrow \Phi(b)$ does not vary with $b \Rightarrow$ (SS2) is violated.

*Proof* Take a non-degenerate baseline distribution $P_B$ (at least two distinct states with positive probability). If (SS2) holds and $H(A) > 0$, there exist $b_1$, $b_2$ with $\Phi(b_1) \neq \Phi(b_2)$ entering $I(B; A)$ with positive probability, so $I(B; A) > 0 \Rightarrow A^* > 0$; the upper bound $A^* \leq 1$ follows from Proposition 2.3. Conversely, $A^* = 0 \Leftrightarrow I(B; A) = 0 \Leftrightarrow$ each $\Phi(b)$ equals the marginal ⇒ all $\Phi(b)$ identical ⇒ contradicts (SS2). If all $\Phi(b) = \delta_{\{a_0\}}$ so that H(A) = 0, this is the fully degenerate limit excluded from the present claim.

**Proposition 2.5** In the deterministic limit, (SS5) is equivalent to strong supervenience; if there exist cross-world base-level isomorphisms preserving the relevant correspondences, global supervenience follows.

*Proof* In the deterministic limit we have $\Phi(b) = \delta_{\{f(b)\}}$, so write $f: B \to A$. Condition (SS5) requires that any $L$-indiscernible $b_1$, $b_2$ induce the same higher-level distribution; since each distribution is a point mass, this is equivalent to $f(b_1) = f(b_2)$. Conversely, if $f$ is constant on every $L$-equivalence class, then the Dirac measures within that class coincide and (SS5) holds. Combining the two directions yields "(SS5) ⇔ $f$ is constant on each $L$-equivalence class," which is the standard formulation of strong supervenience (cf. Kim 1990).

Now assume that any two $L$-worlds admit a base-level isomorphism preserving $f$. If higher-level facts were to differ between these worlds, the values of f would fail to be preserved under the isomorphism—a contradiction. Hence higher-level facts covary with the base structure across all $L$-worlds, giving global supervenience (cf. Kim 1987).

Note. When $\Phi$ is not degenerate, (SS5) constrains an entire family of conditional distributions rather than single point values. The deterministic case is the limiting situation in which every $\Phi(b)$ is Dirac; the probabilistic version is therefore a strict extension of the classical (deterministic) supervenience framework (with (SS3) vacuous in this all-Dirac boundary).

**Proposition 2.6** Let $\pi: B \to C$ be a measurable surjection, and let $\Phi_C(c) = \sum_{\{b:\pi(b)=c\}} \mu_c(b)\,\Phi(b)$, where $\mu_c(b) = P(b \mid c)$. Assume the induced coarse-grained kernel is non-degenerate: equivalently $I(C; A) > 0$ (i.e., there exist $c_1 \neq c_2$ with $\Phi_C(c_1) \neq \Phi_C(c_2)$). If $\Phi$ satisfies (SS1)–(SS5), then $\Phi_C$ also satisfies (SS1)–(SS5) and preserves non-degeneracy (SS2).

*Proof* (SS1) A convex (mixture) combination preserves the law-form invariance. (SS2) By assumption of non-degeneracy ($I(C; A) > 0$), there exist $c_1 \neq c_2$ with $\Phi_C(c_1) \neq \Phi_C(c_2)$; thus (SS2) holds for $\Phi_C$. (SS3) If some $\Phi(b)$ is non-Dirac, $\pi$ compresses information ($I(C; A) \leq I(B; A)$) so there is no law-like bijection back; in the all-Dirac limit (SS3) is vacuous. (SS4) Refinement predicates apply only at the $B$ level; subsequent mixing does not force all outputs to become Dirac. (SS5) $b_1 \sim_L b_2 \Rightarrow \Phi(b_1) = \Phi(b_2) \Rightarrow$ after embedding in their $c$-class the mixture remains class-constant.

**Proposition 2.7** Let $\Phi_1: B \to \Delta(A)$ and $\Phi_2: A \to \Delta(C)$ each satisfy (SS1)–(SS5) and (R)[8], and define $\Psi(b) = \int_A \Phi_2(a)\,d\Phi_1(b)(a)$. Then $\Psi$ also satisfies (SS1)–(SS5).

*Proof* (SS1) Each kernel preserves law-form invariance; integration preserves it. (SS2) Suppose $\Psi(b_1) = \Psi(b_2)$ while $\Phi_1(b_1) \neq \Phi_1(b_2)$. Then $0 = \int_A \Phi_2(a)\,d(\Phi_1(b_1) - \Phi_1(b_2))(a)$, contradicting mixture-separation (R). Hence $\Phi_1(b_1) = \Phi_1(b_2)$, impossible under (SS2) for $\Phi_1$. Thus $\Psi$ is non-degenerate. (SS3) If there exists $b$ with $\Phi_1(b)$ non-Dirac or $a$ with $\Phi_2(a)$ non-Dirac, the composition is information-losing and no law-like bijection back to $B$ exists; if both kernels are everywhere Dirac, (SS3) is vacuous. (SS4) Refining basal predicates only alters indexing granularity for $\Phi_1$; it does not force all composite outputs to become Dirac. (SS5) $b_1 \sim_L b_2 \Rightarrow \Phi_1(b_1) = \Phi_1(b_2) \Rightarrow \Psi(b_1) = \Psi(b_2)$.

---

[8] Regularity (R). For **Proposition 2.7** we impose a mild mixture-separating condition (Fukumizu et al., 2008; Teicher, 1961) on $\Phi_2$: whenever $\mu_1 \neq \mu_2$ are probability measures supported in the convex hull of supports actually used by $\Phi_1(b)$, we have $\int \Phi_2(a)\,d\mu_1(a) \neq \int \Phi_2(a)\,d\mu_2(a)$; constant or affine-collapsed kernels violate this.

**Proposition 2.8** Suppose there exists a legitimate operation $\sigma$ (legitimacy: preserves the given set of basal invariants while only altering mechanism parameters; **Appendix B.7**) producing $\Phi'$ from $\Phi$, and there exists $b \in B$ or a subset $S \subseteq B$ with $d(\Phi(b), \Phi'(b)) \geq \varepsilon > 0$ (or a weighted average difference over $S \geq \varepsilon$), and $\Phi'$ still satisfies (SS1)–(SS5). Then the higher-level structure (or its parametrization $M$) has independent causal discriminability with respect to $A$.

*Proof* Operation $\sigma$ implements $do(\sigma)$. The threshold $\varepsilon$ implies, by the statistical procedures of **Appendix B.4**, rejection of $H_0: \Phi' = \Phi$. Preserved invariants ensure the change is attributable to parameter variation ($\theta \to \theta'$). Thus $do(\sigma)$ alters the conditional distribution family $P(A \mid B)$; the effect of the mechanism parameters on $A$ is both operationalizable and statistically detectable.

**Proposition 2.9** Assume $|A| < \infty$. Let two realization clusters $B_i$, $B_j$ have representative measures $\Phi_i$, $\Phi_j$ satisfying $Sim(B_i, B_j) = 1 - d_{JS}(\Phi_i, \Phi_j) \geq \tau_{high}$, and $TailDiv_{\{ij\}} \leq \kappa$. Then there exists a continuous function $\eta(1 - \tau_{high}, \kappa, |A|)$ with $\eta \to 0$ as $(1 - \tau_{high}, \kappa) \to (0,0)$, such that $|H(\Phi_i) - H(\Phi_j)| \leq \eta(1 - \tau_{high}, \kappa, |A|)$.

*Proof* Since $|A| < \infty$ (assumed), the similarity constraint implies $d_{JS}(\Phi_i, \Phi_j) = 1 - Sim(B_i, B_j) \leq 1 - \tau_{high}$. Standard inequalities (e.g. relations between Jensen–Shannon divergence and total variation, plus Pinsker-type bounds; Cover and Thomas 2005b) then give an upper bound on $d_{TV}(\Phi_i, \Phi_j)$ in terms of $1 - \tau_{high}$. On a finite alphabet, Shannon entropy is uniformly continuous in total variation distance (Fannes / Audenaert–type bounds; Cover and Thomas 2005a), yielding a continuity modulus that goes to 0 as $d_{TV} \to 0$, thus as $1 - \tau_{high} \to 0$. The constraint $TailDiv_{\{ij\}} \leq \kappa$ bounds any additional discrepancy attributed to designated tail mass, so its contribution can be added to the continuity modulus. Collecting these yields a continuous function $\eta(1 - \tau_{high}, \kappa, |A|)$ with $\eta \to 0$ jointly as $(1 - \tau_{high}, \kappa) \to (0,0)$, establishing $|H(\Phi_i) - H(\Phi_j)| \leq \eta(1 - \tau_{high}, \kappa, |A|)$.

**Proposition 2.10** Let $\alpha, \beta \in (0,1)$. $q_\alpha(D_{set})$ be the empirical $\alpha$-quantile of the divergence spectrum $D_{set}$, and $q_{\{1-\beta\}\{perm\}}$ the $(1 - \beta)$-quantile of the permutation (null) distribution (**Appendix B.4**). Suppose there exists $\varepsilon_0 > 0$ such that $q_\alpha(D_{set}) \geq q_{\{1-\beta\}\{perm\}} + \varepsilon_0$, and that $0 < A^* < 1$. Then the pure micro-noise null is rejected at significance level $\beta$, and there exist non-redundant structural distributional differences (an autonomy signal).

***Proof*** The permutation null removes systematic structure while preserving marginals and sample size. Observing the empirical $\alpha$-quantile at least $\varepsilon_0$ above the $(1-\beta)$-quantile of the null has tail probability at most $\beta$ under the null, so the null is rejected. The constraint $0 < A^* < 1$ excludes both degenerate extremes (complete independence and full determinism); the excess divergence is thus attributed to genuine non-redundant structural patterns.

## 3. Discussion

### 3.1 Philosophical Implications

The framework of stochastic supervenience reorients debates about inter-level dependence by identifying the proper relata of law-like constraint: not point-values, but stable families of probability distributions. This shift does more than merely accommodate the "structured uncertainty" found in quantum mechanics, ecology, or complex systems. It provides a methodological bridge on which information-theoretic quantities—such as normalized mutual information—serve as rigorous proxies for metaphysical dependence. By treating the strength of statistical constraint as an objective feature of the system, we can address several persistent problems in the general philosophy of science without inflating our ontology.

Foremost among these is the tension between inter-theoretic dependence and the causal autonomy of the special sciences. A long-standing challenge, often formulated as a generalized exclusion problem (cf. List & Menzies, 2009; Raatikainen, 2010), suggests that if a micro-physical base is sufficient for an outcome, then any macro-scientific model is causally otiose. This objection tacitly assumes that "sufficiency" at the micro-level entails optimal explanatory or causal description. By integrating List and Menzies' (2009) criterion of causal proportionality—which holds that a genuine cause should be specific enough to make a difference without carrying explanatorily irrelevant detail—stochastic supervenience clarifies why macro-models remain indispensable.

In particular, the basal states $B$ strictly fix the macro-level probability distribution $\Phi(b)$, thereby satisfying metaphysical determination, but it typically does so with excessive informational granularity that obscures the relevant causal levers. The effective information increment $\Delta EI > 0$, as defined in **2.1-2.4**, serves as a quantitative detector of proportionality: it identifies scales of coarse-graining at which the causal structure is maximally "sharp," in the sense that interventions over suitably aggregated variables discriminate downstream responses better than do interventions over the full micro-state space (Hoel et al. 2013; Hoel 2017). In these cases, the compressible information in higher-level profiles marks the level at which the system exhibits the best trade-off between specificity and generality (akin to Dennett's "real patterns"; cf. Dennett, 1991; Ladyman et al., 2007). This vindicates macro-autonomy not by denying physical determination, but by showing that macro-descriptions can have superior explanatory depth and causal proportionality.

This perspective also refines our understanding of scientific taxonomy and multiple realizability. Classical discussions (e.g., Fodor, 1974; Putnam, 1967) often treat realization as a binary or categorical relation: either heterogeneous microstructures realize the same macro-property or they do not. The stochastic framework instead reveals a spectrum of distributional similarity. This resonates with Batterman's (2000, 2001) analysis of universality classes: genuine macro-kinds are individuated not merely by pointwise outputs, but by patterns of stability under perturbation and by asymptotic behavior. In our setting, robust realization clusters are precisely those for which the entire distributional shape—captured by our similarity measure $Sim$ and, where relevant, by the tail divergence metric $TailDiv$ introduced in **2.3**—remains law-like under admissible micro-variation. This explains why distinct computational architectures or physical systems may be treated as "functionally equivalent" in standard regimes yet diverge as distinct kinds in high-risk or tail-sensitive regimes: they belong to different probabilistic universality classes despite superficial agreement on typical cases.

Furthermore, the framework clarifies the status of macro-causal variables within mechanistic explanations. Woodward (2004a, 2004c) argues that legitimate macro-variables must support relatively invariant generalizations under interventions. Stochastic supervenience grounds this invariance in the law-governed probability family itself. To say that a macro-variable causes an effect is, on this picture, to say that interventions on that variable systematically alter the parameters of the relevant conditional distributions—shifting means, tightening variances, or reshaping tails—in ways that are not cleanly decomposable into independent contributions from micro-variables. The presence of synergistic information in a PID of $I(Cause; Effect)$ marks scales at which these distributional shifts emerge as irreducible features of the system's organization, rather than as mere aggregates of micro-level influences.

Finally, stochastic supervenience charts a middle path between thoroughgoing determinism and strong emergence, grounded in nomological indeterminacy. In domains ranging from statistical mechanics to generative modelling, the laws do not fix a single trajectory or point-valued outcome, but rather a structured family of probability measures over possible outcomes—a "modal profile" codified by $\Phi$. This picture rejects "extra-lawful forces" while acknowledging that what many special-science explananda latch onto is precisely the shape of this law-governed probability space: the way chances are distributed, not just which outcomes are admissible. This aligns stochastic supervenience with foundational work in the philosophy of physics (e.g., Callender, 2004; Frigg & Werndl, 2021b, 2021a; Loewer, 2001; Myrvold, 2021; Strevens, 2011), on which objective probabilities are genuine structural features of a system's dependence on its base, and not mere summaries of epistemic ignorance.

**3.2 Methodological Justification**

To prevent "higher-level autonomy" and "causal increment" from collapsing into mere verbal re-description, a complementary suite of indicators is required to rule out the suspicions that what we see is "just noise" or "just a trivial bottom-level concatenation." A minimal yet mutually reinforcing bundle is adopted here:

- $A^*$ furnishes a global scale of how strongly the base level constrains the macro distribution (only when it lies strictly between 0 and 1 is there room for substantive discussion);
- $D_{set}$ avoids collapsing everything into a single average distance; it looks at the quantile structure of pairwise divergences, indicating whether there is stable internal shape diversity.
- $TailDiv$ prevents similarity in the core (high probability mass) from masking branching in the shared tails;
- $\Delta EI$, together with PID synergy, shows that the effect emerging after macro compression is not reproducible by a pointwise additive summation of micro variables;
- Tail-weighted statistics (TS, **Appendix B.8**) further re-weights pairs whose central regions are already highly similar, concentrating attention on latent tail discrepancies and flagging pseudo-equivalence or fragility.

Each quantity in isolation is vulnerable to misinterpretation: a high $A^*$ alone might be inflated by a few concentrated peaks; elevated $D_{set}$ quantiles alone could reflect clustered noise; $\Delta EI > 0$ without a substantive synergistic component might be mere redundancy rearranged; pronounced tail divergence with overall constraint near zero may reduce to sporadic sparse events. Juxtaposing them is designed to make these "camouflage pathways" cancel one another's misleading implications.

Their selection follows a complementarity rationale:

- $A^*$ for constraint magnitude;

- $D_{set}$ for structured residual variety;
- $TailDiv$/TS for low-mass fragility;
- $\Delta EI$/PID synergy for intervention-level integrative constraint;
- permutation baselines for falsifiability.

Proposition 2.10 supplies a clean falsification gate: if the upper quantiles of the empirical divergence spectrum decisively exceed the corresponding upper quantiles of the permutation baseline, and $A^*$ lies in (0,1), then a "pure micro-noise" explanation can be rejected—what we observe is not accidental jitter but structured residual variation that the base constraints permit without fully absorbing. TS then augments this by capturing the latent class of cases where "surface similarity conceals tail fissures."

**3.3 Possible Objections and Replies**

**Objection 3.1** Stochastic supervenience is merely an epistemic-ignorance repackaging: with sufficiently fine-grained base information, the probabilities would collapse into determinate values.

***Reply*** Stochastic supervenience blocks a purely "epistemic refinement → Dirac $\delta$-collapse" route at the structural level through (SS2) and (SS4). (SS2) requires the presence of genuine distributional differences; (SS4) states that after any legitimate refinement of base predicates (holding the law set $L$ fixed), if $\Phi$ remains non-Dirac (i.e., does not collapse to Dirac $\delta$ point masses) and the distance structure (e.g., $d_{JS}$, the divergence spectrum $D_{set}$, and $TailDiv$) does not collapse, then the residual randomness is attributable to the pattern of laws rather than a cognitive deficit.

Criterion: if, as one successively introduces additional micro-level distinctions (finer parameters, local state labels), $A^*$ together with representative pairwise distances remains stable (neither drifting monotonically toward 1 nor evaporating toward 0), the system exhibits a law-locked family of distributions instead of a progressively vanishing layer of subjective uncertainty. Paradigmatic cases include quantum Born statistics, ecological power-law regularities, and the stable profile by which a trained deep generative model reproduces empirical frequencies on independent test sets; these "frequency laws" retain their form when further observational channels are added. Conversely, if under refinement $\Phi(b)$ incrementally $\delta$-izes (approaches Dirac $\delta$ measures), $A^*$ monotonically tends toward 0 or 1, and $D_{set}$ plus $TailDiv$ collapse to permutation/noise baselines, the randomness is traced to epistemically unresolved variables, and stochastic supervenience makes no further ontological claim.

**Objection 3.2** The metaphysical base relation (e.g. grounding (Schaffer, 2017)) requires determinate functional dependence; probabilistic structure would erode explanatory directionality.

***Reply*** Stochastic supervenience is meant to model some central aspects of grounding-like base relations in probabilistic settings. The base structure together with $L$ law-fixes $\Phi$ (SS1 with SS5); the members of $\Phi$ form a minimally sufficient family (no further unification is possible without enlarging the base's referential resources), and, whenever $\Phi$ contains at least one non-Dirac conditional, there is no symmetric reverse law-like reconstruction (SS3). Explanatory directionality is secured by the fact that any systematic shift in the higher-level distribution—or in its parameters—can be traced to an admissible perturbation path of variables in $B$ and their mechanistic parameters (Proposition 2.8), without invoking independent higher-level primitive causes. Compared with classical functional determination, distributional entailment simply replaces value-locking with measure-family locking, while retaining metaphysical asymmetry and a corresponding form of explanatory minimality.

**Objection 3.3** Stochastic supervenience can be trivially degenerated: any deterministic function $f: B \to A$ can be recast as $f: B \to \Delta(A)$ (assigning a Dirac measure), thereby draining the concept of its discriminatory force.

***Reply*** The discriminative force of stochastic supervenience lies precisely in its explicit archiving of the degenerate case. The deterministic limit (Proposition 2.3) is merely a boundary point of the $\Phi$-space, whereas genuine stochastic supervenience requires: (1) non-degeneracy (SS2); (2) $A^* \in (0,1)$ (Proposition 2.4); (3) a divergence spectrum exhibiting statistically significant structure (Proposition 2.10); (4) at least some macro-aggregates for which $\Delta EI > 0$ or the synergy share increases. If all $\Phi(b)$ are Dirac (or mutually identical), these criteria fail systematically, and the model is explicitly tagged as deterministic limit / degenerate rather than conflated with the non-degenerate case. The concept is thus not emptied; its discriminative power is preserved via a boundary-marking strategy.

**Objection 3.4** Stochastic supervenience merely elevates familiar statistical fitting into metaphysical rhetoric; it supplies no incremental standards of prediction or explanation.

***Reply*** Conventional statistical models can fit the same marginal and conditional probabilities, but they do not posit an ontological structure linking law-like fixation to hierarchical differentiation. Stochastic supervenience proceeds by:

(1) law-like fixation (SS1) together with distributional necessity (SS5), treating the shape of the distribution as itself subject to nomic constraint;

(2) non-degeneracy and divergence-spectrum significance (Proposition 2.10) to discriminate mere noise perturbations from structural variation;

(3) $\Delta EI$ and the synergy increment to determine whether the intervention-sensitive structure revealed by macro compression is irreducibly additional, rather than reconstructible from a small set of low-complexity local functions;

(4) tail-sensitive indicators ($TailDiv$, TS) that detect latent bifurcations concealed under ordinary mean/variance approximations.

The predictive increment arises when two models coincide on point estimates or the first few moments, yet yield empirically discriminable forecasts regarding tail-event frequencies, the conditional mutual information network (testable via visualizing estimated $I(X; A \mid Y)$), or coarse-grained intervention responses (e.g. enhanced robustness or divergent risk tails). These divergences permit empirical adjudication of metaphysical sufficiency—namely, which model actually fixes the global structure of $\Phi$.

**Objection 3.5** Composite charge of "arbitrariness," "circularity," and "vacuity": the choice of divergence measures, thresholds, and coarse-grainings is arbitrary; the appeal to information-theoretic quantities is circular (it presupposes dependence to measure dependence); and the resulting notion of stochastic supervenience risks vacuity.

*Reply* The framework is not committed to any unique metric or threshold choice. Instead, it specifies a family of admissible indices satisfying modest formal properties (e.g. separation, triangle inequality) and checks robustness of qualitative conclusions across reasonable variants (e.g. J-S vs Wasserstein; different $p_0$, $\tau_{high}$). Circularity worries are diffused by the fact that mutual information and divergence measures can be estimated from data independently of any prior metaphysical commitments; what the framework adds is an interpretation: ranges of these quantities are used as operational criteria for when higher-level structure is non-degenerate, non-noise, and intervention-salient.

Vacuity is avoided by explicit failure conditions. A system fails to instantiate non-degenerate stochastic supervenience whenever:

- (SS2) collapses (all $\Phi(b)$ identical),
- or $A^*$ tends to 0 or 1 under refinement,
- or $D_{set}$ and $TailDiv$ fail to exceed permutation baselines,

- or $\Delta EI \leq 0$ for all admissible coarse-grainings.

In such cases, the framework simply records that there is no substantive higher-level probabilistic structure over and above what is already encoded locally at the base; the supervenience relation collapses back to the deterministic or trivial limit. Stochastic supervenience is thus empirically defeasible: it earns its keep only where its indices pick out distinctive, law-locked distributional patterns.

## 4 Conclusion

The project of stochastic supervenience begins from an apparently modest adjustment: the determinandum of supervenience is shifted from higher-level point values to higher-level probability distributions. Yet once this shift is made explicit and axiomatized, several payoffs follow.

Formally, the framework treats kernels $\Phi: B \to \Delta(A)$ as the primary vehicles of dependence. The axioms SS1–SS5 secure law-like fixation, non-degeneracy, directional asymmetry, epistemic closure, and distributional necessitation. Classical deterministic supervenience is shown to be a boundary case in which every $\Phi(b)$ is Dirac; strong and global supervenience re-emerge as specializations of (SS5) under Dirac limits. Coarse-graining and composition preserve the structure of stochastic supervenience under mild conditions, ensuring that inter-level dependence behaves well under standard scientific idealizations.

Methodologically, information-theoretic and divergence-based indices provide operational diagnostics. They allow us to:

- quantify the strength of law-like constraint (via $A^*$), and
- chart residual distributional variety (via $D_{set}$),
- distinguish robust from fragile or pseudo-equivalent realizations (via $Sim$ and $TailDiv$),
- and detect macro-level gains in intervention-sensitive structure (via $\Delta EI$ and synergy).

These diagnostics make it possible to reject "pure noise" explanations in cases where higher-level probabilistic structure is genuinely law-locked and experimentally discriminable.

Philosophically, stochastic supervenience offers a unified language for describing dependence in domains where structured uncertainty is central: quantum measurement, statistical mechanics, ecological scaling laws, complex systems, and data-driven generative models. It preserves the asymmetry and priority of the base level while acknowledging that what laws fix, in many scientifically central cases, is a family of chances rather than a single outcome.

In this sense, stochastic supervenience is both a conservative and an expansive move. It is conservative in that it preserves all classical results as a Dirac boundary; it is expansive in that it opens a structured space between strict determinism and unregulated randomness, populated by law-governed probability profiles which can be probed, compared, and tested.

The broader significance of this shift is twofold. First, it aligns metaphysical accounts of dependence more closely with actual scientific practice in stochastic and statistical domains. Second, it provides a platform for future work on counterfactual semantics and evaluative criteria in probabilistic settings, where questions of autonomy, emergence, and inter-theoretic reduction can be addressed with both conceptual and quantitative tools.

**Appendix A Illustrative Cases**

**A.0 Common Setup and Thresholds**

Similarity: $Sim(B_i, B_j) = 1 - d_{JS}(\Phi_i, \Phi_j)$, where $d_{JS}$ is the (square-root) J–S distance taking values in $[0,1]$.

Tail set: Let $p_0 = 0.05$. If $\Phi_k(a) < p_0$, then $a$ is treated as a "rare" event for cluster $B_k$. The shared tail $S_{\{ij\}}$ is the set of symbols whose probabilities are below $p_0$ in both clusters.

$TailDiv$: Compute the total variation distance $d_{TV}$ between the two clusters after restricting to $S_{\{ij\}}$ and renormalizing the probabilities on that tail subset.

Illustrative thresholds (tunable but fixed ex ante): $\tau_{high} = 0.85$, $\tau_{low} = 0.75$, $\theta_{tail} = 0.15$, $\theta_{pseudo} = 0.35$. Thresholds should be chosen before hypothesis testing and, where possible, accompanied by sensitivity analyses. Classification rules (see **2.3**).

Note: In Case A.1 the salient differences do not lie in the tail (most differing categories exceed $p_0$), so $TailDiv$ is mainly used in A.2; in A.1 we still compute $Sim$ to show how "body mass redistribution" drives stratification.

**A.1 Neural Network Classification**

**A.1.1 Basal States and Output Distributions**

This toy example mirrors standard supervised classification settings in machine learning, where different architectures realize similar input–output mappings while exhibiting distinct conditional output distributions under degraded inputs. Basal state $b$ is given by (network architecture $N_k$, input clarity $q \in \{cl, bl\}$). Output alphabet $A = \{cat, dog, unsure\}$:

$$\Phi(b_{\{N1\}^{\{cl\}}}) = (0.95, 0.03, 0.02)$$

$$\Phi(b_{\{N1\}\{bl\}}) = (0.60, 0.30, 0.10)$$

$$\Phi(b_{\{N2\}\{cl\}}) = (0.94, 0.04, 0.02)$$

$$\Phi(b_{\{N2\}\{bl\}}) = (0.42, 0.40, 0.18)$$

With clear inputs (cl) the two networks yield almost identical distributions; with blurred inputs (bl) there is marked redistribution in the body probabilities (cat decreases; dog and unsure increase), exhibiting structured uncertainty rather than independent noise.

**A.1.2 Realization Clusters, Representative Measures, and Similarity**

Define realization clusters $B_{\{N1\}} = \{b_{\{N1\}\{cl\}}, b_{\{N1\}\{bl\}}\}$ and $B_{\{N2\}} = \{b_{\{N2\}\{cl\}}, b_{\{N2\}\{bl\}}\}$; for comparison we also examine pairs at a fixed clarity level. Uniform weights are used.

Example J–S distances and similarities (approximate):

$$d_{JS}(\Phi(b_{\{N1\}\{cl\}}), \Phi(b_{\{N2\}\{cl\}})) \approx 0.023 \Rightarrow Sim \approx 0.977 \text{ (high body agreement)}$$

$$d_{JS}(\Phi(b_{\{N1\}\{bl\}}), \Phi(b_{\{N2\}\{bl\}})) \approx 0.159 \Rightarrow Sim \approx 0.841 \text{ (body redistribution increases divergence)}$$

At $p_0 = 0.05$, (dog, unsure) are partly still "low probability" under cl but both $\geq p_0$ under bl. Thus, the divergence here is not categorized as tail-structure difference but as reallocation within the high-probability body. Classification:

(cl vs cl): $Sim \geq \tau_{high}$ and no material tail difference → Robust

(bl vs bl): $\tau_{low} \leq Sim < \tau_{high}$ → Fragile

Hence, if one aggregated cl and bl within each network into a single representative $\Phi_{\{N1\}}$, $\Phi_{\{N2\}}$, the source of fragility (conditional scenario shift) would be partially obscured, indicating a need for scenario stratification or conditional analysis.

**A.2 Multi-Cluster Markov System**

**A.2.1 Basal States and Output Distributions**

Basal state set (illustrative) $\mathcal{B} = \{s1, s2, f1, f2, p1, p2, r1, r2\}$. Via a coarse-graining rule the output alphabet is $A = \{C1, C2, C3, r1, r2\}$, where $C1 = \{s1, s2\}$, $C2 = \{f1, f2\}$, $C3 = \{p1, p2\}$; $r1$, $r2$ are retained as rare event symbols. Single-step output distributions (order: $C1, C2, C3, r1, r2$):

$$s1: (0.47, 0.43, 0.06, 0.02, 0.02)$$

$$s2: (0.45, 0.45, 0.06, 0.02, 0.02)$$

$$f1: (0.46, 0.41, 0.09, 0.02, 0.02)$$

$$f2: (0.44, 0.39, 0.09, 0.06, 0.02)$$

$$p1: (0.47, 0.42, 0.06, 0.045, 0.005)$$

$$p2: (0.45, 0.44, 0.06, 0.045, 0.005)$$

The body ($C1, C2, C3$) is highly aligned; differences concentrate in the ($r1, r2$) structure.

### A.2.2 Realization Clusters and Representative Measures

Using uniform weights:

$$B_R = \{s1, s2\} \rightarrow \Phi_R = (0.46, 0.44, 0.06, 0.02, 0.02)$$

$$B_F = \{f1, f2\} \rightarrow \Phi_F = (0.45, 0.40, 0.09, 0.04, 0.02)$$

$$B_P = \{p1, p2\} \rightarrow \Phi_P = (0.46, 0.43, 0.06, 0.045, 0.005)$$

### A.2.3 Body Similarity ($Sim$)

Approximate pairwise distances:

$$d_{JS}(\Phi_R, \Phi_F) \approx 0.073 \Rightarrow Sim \approx 0.927$$

$$d_{JS}(\Phi_R, \Phi_P) \approx 0.084 \Rightarrow Sim \approx 0.916$$

$$d_{JS}(\Phi_F, \Phi_P) \approx 0.078 \Rightarrow Sim \approx 0.922$$

All bodies meet the $\tau_{high} = 0.85$ threshold, indicating macro-level body resemblance.

### A.2.4 Tail Structure and $TailDiv$

With $p_0 = 0.05$, $r1$ and $r2$ enter the shared tail. After tail renormalization:

$$B_R: (0.02, 0.02) / 0.04 \rightarrow (0.50, 0.50)$$

$$B_F: (0.04, 0.02) / 0.06 \rightarrow (0.667, 0.333)$$

$$B_P: (0.045, 0.005) / 0.05 \rightarrow (0.90, 0.10)$$

$TailDiv$ (total variation distances):

$$TailDiv(R, F) = 0.167$$

$$TailDiv(R, P) = 0.400$$

$$TailDiv(F, P) = 0.233$$

### A.2.5 Stratification Results

$$B_R \text{ vs } B_F: Sim \geq \tau_{high} \text{ and } \theta_{tail} < 0.167 < \theta_{pseudo} \rightarrow \text{Fragile}$$

$$B_R \text{ vs } B_P: Sim \geq \tau_{high} \text{ and } TailDiv = 0.400 > \theta_{pseudo} \rightarrow \text{Pseudo-equivalent}$$

$$B_F \text{ vs } B_P: Sim \geq \tau_{high} \text{ and } 0.15 < 0.233 < 0.35 \rightarrow \text{Fragile}$$

### A.2.6 Interpretation

If one considered only the body (ignoring tail-renormalized divergence), the three clusters would appear "nearly equivalent." Incorporating $TailDiv$ separates moderate proportional shifts (Fragile) from pronounced skew (Pseudo-equivalent). Pseudo-equivalence captures cases where body alignment masks substantial differences in functional or risk profiles; overlooking this would exaggerate true redundancy at the realization level.

**A.2.7 Method Portability**

In empirical settings one needs only:

(1) Estimate $\Phi(b)$;

(2) Choose $p_0$ and distances $d_{JS}$ / $d_{TV}$;

(3) Form candidate clusters $B_i$ and compute $\Phi_i$;

(4) Compute $Sim$ and $TailDiv$;

(5) Apply classification thresholds;

(6) Use bootstrap to obtain confidence intervals for $Sim$ / $TailDiv$, assigning a label only when the interval lies stably within a category.

**A.3 Summary**

Case A.1 illustrates fragility driven by conditional body redistribution; Case A.2 shows fragility and pseudo-equivalence driven by tail proportion shapes. Together they demonstrate that distribution-level multiple realizability requires joint inspection of body and tail; reliance solely on aggregated or body-only metrics risks misclassifying structural differences across realization layers.

**Appendix B Metrics and Statistical Considerations**

A uniform Dirichlet / Laplace smoothing scheme is applied throughout (see Krichevsky and Trofimov 1981). Report point estimates together with bootstrap or permutation 95% confidence intervals / p-values (Efron and Tibshirani 1994, 153-55). Notation: $A^* = I(B; A)/H(A)$; $\Delta EI = EI_{macro} - EI_{micro}$.

**B.1 Estimating $A^*$**

Use the plug-in estimator with Dirichlet ($\alpha$) smoothing ($\alpha = 1/K$ or $1/2$) (Efron and Tibshirani 1994; Paninski 2003). For small samples apply the Miller–Madow correction to mutual information bias (Panzeri and Treves 1996). Bootstrap confidence intervals are recommended.

**B.2 J–S and Tail Divergence**

In multi-class settings entropy bias under maximum likelihood is $O(1/n)$ (see Paninski 2003); avoid zeros by using the same smoothing as in B.1. $TailDiv$: fix a threshold $p_0$ (do not tune repeatedly on the data), form the shared tail $S_{\{ij\}}$, renormalize the tail distributions, then compute $d_{TV}$. For high body-similarity pairs, $TailDiv$ values can be aggregated (weighted average or quantiles) to yield a tail sensitivity summary distinguishing robust / fragile / pseudo-equivalent cases. If the shared tail mass is 0, we set $TailDiv = 0$ as a conservative convention—interpreted as "no comparable tail," not "identical tail shape." Large body differences remain captured by $Sim$.

### B.3 Computing $\Delta EI$

Estimate interventional distributions $p_{do}$ from a causal graph (see Pearl 2009); use a matched intervention scheme at micro and macro-levels (often uniform). If $\Delta EI > 0$ and its confidence interval excludes 0 (or Bayes factors support it; see Kass and Raftery 1995), treat macro aggregation as providing causal information gain (Hoel et al. 2013).

### B.4 Permutation Tests

Randomly permute basal labels or cluster labels M times and recompute the divergence spectrum / $TailDiv$ (see Good 2000b). Compute the p-value as $p = (1 + \#\{stat_{perm} \geq stat_{obs}\}) / (M + 1)$. For multiple indices / quantiles apply FDR control (Benjamini and Hochberg 1995).

### B.5 Parameter Regularization

Minimum description length: $MDL(\Phi) = L_{model} - \log p(data \mid \Phi)$ (Rissanen 1978), where $L_{model}$ is the model codelength (distinct from the law set $L$). We approximate $L_{model}$ via the BIC term $0.5 \cdot df \cdot \log n$ (see Schwarz 1978) or use a predefined code length, select the model minimizing $MDL(\Phi)$, and report $\Delta MDL$ to the next-best alternative.

### B.6 Cross-Context Robustness

Given $\Phi_k$ estimated on distinct data domains $D_k$, if the global mean (or median) of $d(\Phi_i(b), \Phi_j(b))$ stays below a threshold (set, e.g., by the 95% resampling quantile within a domain; Efron and Tibshirani 1994, 152-65) and entropy, $TailDiv$, and related structural indices remain within preset tolerance bands, declare structural robustness.

### B.7 Operationalizing Intervention Legitimacy

Require that a set of basal invariants $I$ (e.g., energy, mean firing rate) show no meaningful change pre/post intervention; use equivalence tests (TOST) or narrow-interval Bayes factors (cf. Schuirmann 1987; Kass and Raftery 1995), not merely $p > 0.05$. Only after output distribution differences are significant (G-test, mutual information permutation test; cf. Good 2000a)) do $\Delta EI$ and PID interpretations carry causal legitimacy.

**B.8 Tail Sensitivity Index (TS)**

Steps: (1) Select the set $P = \{(i,j) \mid Sim(B_i, B_j) \geq \tau_{high}\}$ of high body similarity cluster pairs.

(2) For each pair compute $TailDiv_{\{ij\}}$ and shared tail mass $Z_{\{ij\}} = \frac{1}{2}(\Phi_i(S_{\{ij\}}) + \Phi_j(S_{\{ij\}}))$.

(3) Define $TS = (\sum_{\{(i,j) \in P\}} TailDiv_{\{ij\}} \cdot Z_{\{ij\}}) / (\sum_{\{(i,j) \in P\}} Z_{\{ij\}})$.

Interpretation: conditional on body similarity, TS measures the average intensity of latent tail divergence. High TS signals pseudo-equivalence or fragility risk. Compare to a permutation baseline to obtain a z-score.

Schwarz, G. (1978). Estimating the Dimension of a Model. *The Annals of Statistics*, *6*(2), 461 ~ 464.

Shagrir, O. (2002). Global Supervenience, Coincident Entities and Anti-Individualism. *Philosophical Studies*, *109*(2), 171 ~ 196. https://doi.org/10.1023/A:1016224703009

Shagrir, O. (2013). Concepts of Supervenience Revisited. *Erkenntnis*, *78*(2), 469 ~ 485. https://doi.org/10.1007/s10670-012-9410-7

Shannon, C. E. (1948). A mathematical theory of communication. *The Bell System Technical Journal*, *27*(3), 379 ~ 423. https://doi.org/10.1002/j.1538-7305.1948.tb01338.x

Sober, E. (1999). The Multiple Realizability Argument Against Reductionism. *Philosophy of Science*, *66*(4), 542 ~ 564. https://doi.org/10.1086/392754

Stenwall, R. (2021). A grounding physicalist solution to the causal exclusion problem. *Synthese*, *198*(12), 11775 ~ 11795. https://doi.org/10.1007/s11229-020-02829-3

Strevens, M. (2011). Probability out of Determinism. In C. Beisbart & S. Hartmann (eds.), *Probabilities in Physics* (pp. 339 ~ 364). Oxford University Press. https://doi.org/10.1093/acprof:oso/9780199577439.003.0013

Teicher, H. (1961). Maximum Likelihood Characterization of Distributions. *The Annals of Mathematical Statistics*, *32*(4), 1214 ~ 1222.

Vapnik, V. N. (1999). An overview of statistical learning theory. *IEEE Transactions on Neural Networks*, *10*(5), 988 ~ 999. https://doi.org/10.1109/72.788640

Villani, C. (2009). The Wasserstein distances. In C. Villani (eds.), *Optimal Transport: Old and New* (pp. 93 ~ 111). Springer. https://doi.org/10.1007/978-3-540-71050-9_6

Vinh, N. X., Epps, J., & Bailey, J. (2010). Information Theoretic Measures for Clusterings Comparison: Variants, Properties, Normalization and Correction for Chance. *Journal of Machine Learning Research*, *11*(95), 2837 ~ 2854.

Wallace, D. (2019). The Necessity of Gibbsian Statistical Mechanics. In *Statistical Mechanics and Scientific Explanation* (pp. 583 ~ 616). WORLD SCIENTIFIC. https://doi.org/10.1142/9789811211720_0015

Wigner, E. P. (1963). The Problem of Measurement. *American Journal of Physics*, *31*(1), 6 ~ 15. https://doi.org/10.1119/1.1969254

Williams, P. L., & Beer, R. D. (2010). *Nonnegative Decomposition of Multivariate Information* (No.
34